# Effects of quantum zero-point spin fluctuations on the ground state of plutonium

P.V. Ratnikov[1] and A. Solontsov[1,2*]

[1]N.L. Dukhov Research Institute for Automatics, Sushchevskaya str., 22, 127055 Moscow, Russia

[2]State Center for Condensed Matter Physics, 6/3 str. M. Zakharova, Moscow 115569, Russia

Dated: October 27, 2013

The magnetic stability of δ-plutonium is analyzed taking into account zero-point spin fluctuations. Within the generalized theory of spin fluctuations described within a simple phenomenological model neglecting its spatial dispersion we show that zero-point local magnetic moments are giant ($\left(M_L^2\right)_{Z.P.} \approx 2\mu_B^2$ / at.Pu) and suppress the magnetic order predicted by *ab initio* calculations resulting in the observed paramagnetic state of δ-plutonium.

## Introduction

In spite of decades of extensive research, investigations of plutonium still remain in the focus of physics of strongly correlates electron systems. Plutonium undergoes a cascade of structural phase transitions resulting in six allotropic phases of which the $\delta$-phase with the fcc lattice is considered to be most important. It is stable above 593K and may be stabilized at lower temperatures by adding small amounts of, e.g., Ga or Al. The mechanism of this stabilization as well as the nature of phase transformations are still unknown.

Numerous band-structure calculations (see the review [1] and references therein) still support the picture suggested by Johansson [2(8)]: plutonium should be viewed as an itinerant electron system with boderline itinerant vs localized $5f$-electrons which can be treated within the *ab initio* density functional theory (DFT). However, most of the DFT calculations predict a magnetic ground state of plutonium though no magnetically ordered state was observed experimentally [1]. Nevertheless, as was emphasized by Sonderlind et al. [3], magnetism plays an essential role in all phases of plutonium.

At least, this should be due to the fact that all phases of plutonium are close to magnetic instabilities and must possess strong spin fluctuations (SF). Up to now SF in plutonium were not observed directly in inelastic neutron scattering experiments [1]. However, there are strong indirect indications of their effects in plutonium beginning from the discovered in 1972 by Arko et al. [4] influence of SF on electroresistivity of plutonium. Later were discussed effects of SF in plutonium on the thermal resistivity, thermoelectric power, magnetic susceptibility, specific heat and NMR relaxation rate, which are generally enhanced by the presence of paramagnons [5]. Recently the effects of SF were shown to successfully explain the anomalies of thermal expansion of $\delta$-Pu due to the strong magnetovolume effect [6].

Among many unresolved problems of plutonium one of the most important is, why DFT calculations do not explain the paramagnetic ground state of plutonium and predict an erroneous magnetically ordered state with the ordered moments in the range 1-2 $\mu_B$ [1]. The answer could also be related to the effects of SF. Besides thermal SF affecting thermal properties of plutonium it possesses zero-point SF existing in the ground state and at elevated temperatures [7]. Zero-point SF usually have an amplitude about one Bohr magneton per atom and are treated as giant [8]. Giant zero-point SF lead to strong spin anharmonicity and cannot be treated within the perturbative approach which found the basis for the conventional SF theory of weakly anharmonic paramagnons [9]. An essential success in description of SF accounting for both giant SF and strong spin anharmonicity was made within the developed in 90-s generalized theory (GT) of SF of Solontsov and Wagner [10]. Unlike the conventional theory of SF [9] it does not use the perturbative approach and is based on a variational procedure. The GT shows that effects of zero-point SF and spin anharmonicity strongly renormalize the ground state and change the criterion of magnetic instability which becomes essentially different from the well-known Frenkel-Stoner one. Zero-point SF and spin anharmonicity were shown to suppress magnetic instabilities with respect to the conventional SF description [9] not accounting for the zero-point effects.

On the other hand, the DFT calculations of the ground state properties of many-electron systems is usually based on the local density approximation (LDA) which do not account for zero-point effects [11]. Recently the LDA version of DFT was incorporated into the GT to analyze the ground state of $\alpha$- and $\delta$-plutonium [11]. Zero-point effects were shown to partly suppress the magnetically ordered state both in $\alpha$- and $\delta$-plutonium which still remain magnetically ordered. This incomlete supression of the ordered ground state is probably due to the approximations made in the paper [11]. Namely, to calculate the dynamical susceptibility and other quantaties entering the GT the authors [11] used random phase (RPA) and LDA approximations and assumed strong spatial dispersion of SF, which may be not justified for plutonium.

In the present paper we analyze the stability of the ground state of plutonium taking into account local SF neglecting their spatial dispersion. Previously this model was successfully used to describe temperature dependencies of resistivity [4] and thermal expansion [10] of the plutonium. To account for the zero-point effects we use the GT approach in the weak spin anharmonicity limit justified by the measured weak temperature dependence of magnetic susceptibility [1].

**The model of spin fluctuations of plutonium.**

We start with the following model for the dynamical magnetic susceptibility $\chi(\mathbf{Q},\omega)$ of plutonium defining the spectrum of SF, where $\mathbf{Q}$ and $\omega$ are the wavevector and frequency of SF. At the moment, when any data on spatial and time dispersions of the dynamical magnetic susceptibility of Pu is lacking it is reasonable to illustrate SF effects using the most simple approximation for $\chi(\mathbf{Q},\omega)$, assuming that its spatial dispersion is relatively weak and neglect its dependence on the wavevector, $\chi(\mathbf{Q},\omega) \approx \chi(\omega)$. This allows to present the spectrum of SF defined by the imaginary part of the dynamical susceptibility for not very high frequencies in the following simple form

$$\operatorname{Im}\chi(\omega) = \chi(T)\frac{\omega\omega_{SF}}{\omega^2 + \omega_{SF}^2}, \tag{1}$$

where $\chi(T)$ is the static magnetic susceptibility of plutonium which we take from the measurements [12], $\hbar\omega_{SF} \approx 0{,}187$ eV is the temperature-independent characteristic frequency of SF inferred from the thermal expansion data [6].

The key parameters of the SF theory are the squared local moments or squared averaged amplitudes of SF, which are given by the fluctuation-dissipation theorem

$$M_L^2 = 12\hbar\int_0^\infty (d\omega/2\pi)\operatorname{Im}\chi(\omega)\left(N_\omega + \frac{1}{2}\right) \equiv \left(M_L^2\right)_{Z.P.} + \left(M_L^2\right)_T, \tag{2}$$

Here the factors $N_\omega = [\exp(\hbar\omega/k_B T) - 1]^{-1}$ and $1/2$ are related to thermal and zero-point SF, respectively, which in our model can be presented explicitly

$$(M_L^2)_{Z.P.} = \frac{3\nu\hbar\omega_{SF}\chi(0)}{2\pi\Omega}\ln[1 + (\frac{\omega_c}{\omega_{SF}})^2],$$

$$(M_L^2)_T = \frac{3k_B T}{\Omega}\nu\chi(0)g(\frac{\hbar\omega_{SF}}{2\pi k_B T}). \tag{3}$$

where $\nu$ and $\Omega$ - are the number of Pu atoms in the unit cell (for the fcc $\delta$-Pu $\nu = 4$) and its volume, $g(y) = 2y[\ln y - 1/2y - \Psi(y)]$, $\Psi(y)$ - is the Euler's psi function. The temperature dependences of local moments (3) in the stabilized $\delta$-Pu are illustrated by Fig. 1.

Within the GT theory the zero-point local moments effects the ground state properties and thermal ones define the temperature dependence of the magnetic susceptibility $\chi(T)$. Namely, the well-known Frenkel-Stoner criterion of the magnetic instability

$$1 + \Psi \chi^{(0)} < 0, \quad (4)$$

where $\chi^{(0)}$ is the Pauli susceptibility and $\Psi$ describes the exchange coupling of electrons in the weak anharmonicity limit is changed for [10]

$$\chi^{-1}(0)\chi^{(0)} = 1 + \Psi \chi^{(0)} + \frac{5}{3}\gamma \chi^{(0)}(M_L^2)_{Z.P.} < 0, \quad (5)$$

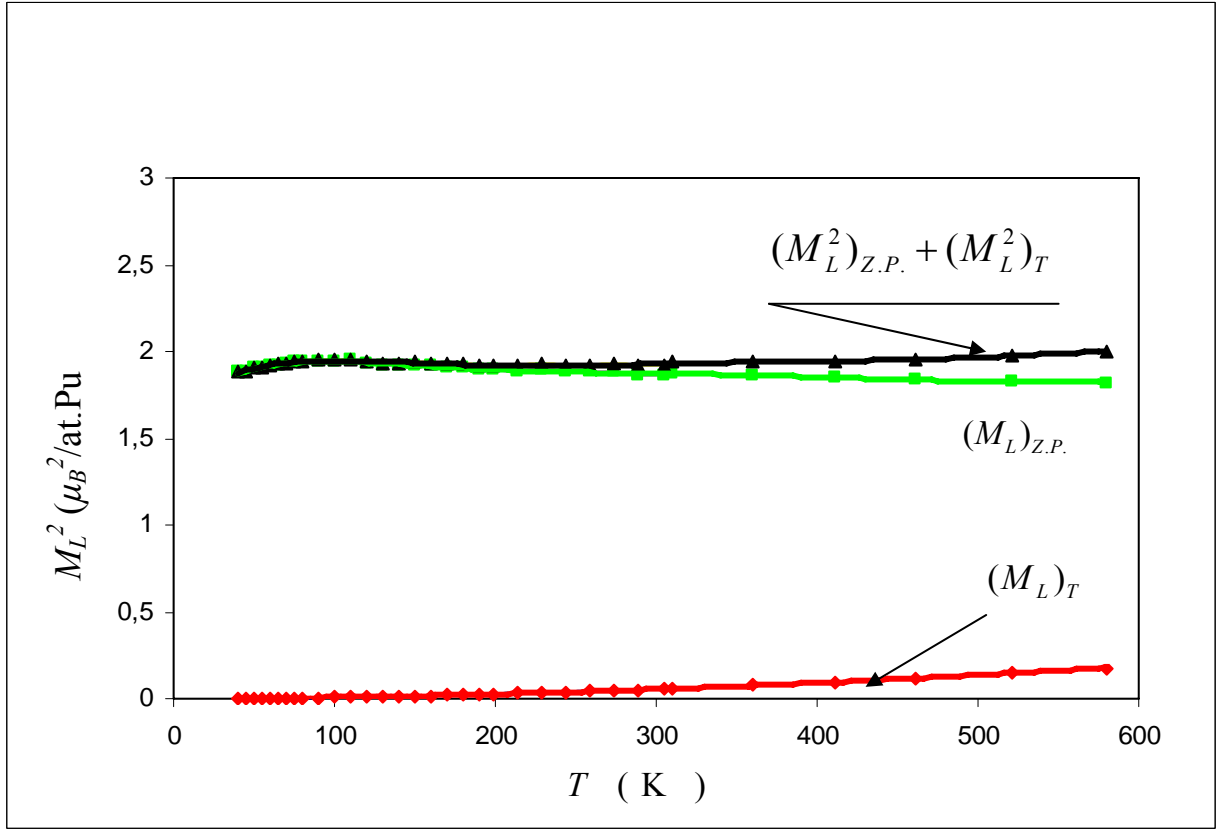

Fig. 1. The temperature dependencies of the squared thermal and zero-point local magnetic moments in the stabilized $\delta$-Pu calculated from (3) with the cut-off energy $\hbar\omega_c \approx 1\,\text{eV}$ and magnetic susceptibility $\chi(0)$ taken from [12].

where $\gamma$ is the positive mode-mode coupling constant of SF. For the temperature-dependent magnetic susceptibility one has [9,10]

$$\chi^{-1}(T) = \chi^{-1}(0)\left[1 + \frac{5}{3}(M_L^2)_T \chi(0)\gamma\right] \quad (6)$$

The temperature dependence of the magnetic susceptibility $\chi(T)$ in $\delta$-Pu described by (6) together with the measured data [12] is shown on Fig. 2.

As one can see from (5) zero-point SF described by the term $\sim (M_L^2)_{Z.P.}$ can suppress magnetic instability arising without account of zero-point SF and lead to a stable paramagnetic phase, where

$$1 + \Psi \chi^{(0)} + \frac{5}{3}\gamma\chi^{(0)}(M_L^2)_{Z.P.} > 0, \qquad (7)$$

provided

$$\frac{5}{3}(M_L^2)_{Z.P.} > |\gamma\chi_0|^{-1}. \qquad (8)$$

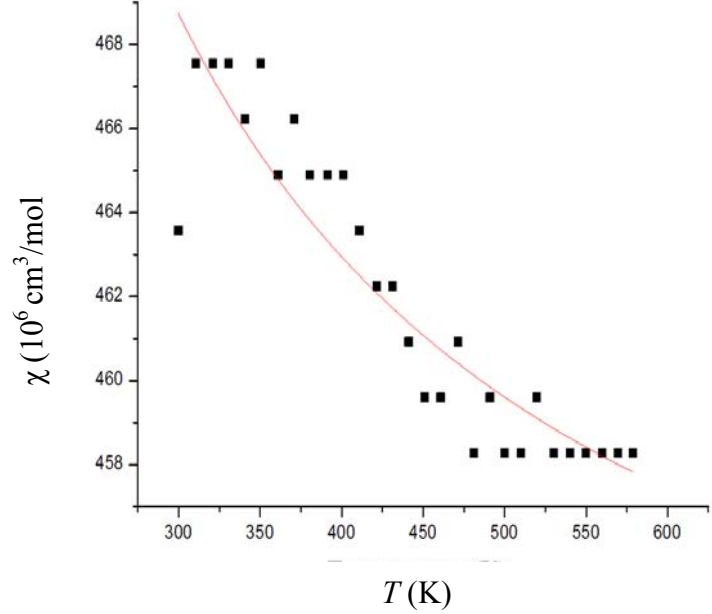

Fig. 2. The temperature dependence of the magnetic susceptibility of $\delta$-Pu calculated from (6) (solid line) and measured values of the susceptibility (black squares) taken from [12].

Here $\chi_0 = \chi^{(0)}/(1+\Psi\chi^{(0)})$ is the paramagnetic susceptibility without account of SF which we assume to be negative in plutonium.

Using measured temperature dependence of the magnetic susceptibility [12] and calculated values of local moments (3) we can show that inequality (6) holds for the $\delta$-plutonium. Taking the measured variation of the susceptibility [12] $\chi(40K) - \chi(575K) \approx 140 \times 10^6$ cm$^3$/mole and the calculated from (6) squared thermal moment $\left(M_L^2\right)_{T=575K} \approx 0.25\mu_B^2$/at.Pu, using Fig. 2 we can estimate the value $(\chi(0)\gamma)^{-1} \approx 2\mu_B^2$/at.Pu and $(\chi_0\gamma)^{-1} \approx -1.25\mu_B^2$/at.Pu. Estimating from (3) the squared zero-point local moment $\left(M_L^2\right)_{Z.P.} \approx 2\mu_B^2$/at.Pu we see that the inequality (7) holds for $\delta$-plutonium which make the initially ordered ground state in the Frenkel-Stoner model paramagnetic with account of zero-point SF.

**Conclusions**

To conclude, we show, that zero-point SF essentially affect the ground state of the stabilized $\delta$-plutonium and suppress the magnetic order predicted by DFT calculations. We assume that $\delta$-plutonium is magnetically ordered provided zero-point effects are neglected. To account for zero-point SF we use a simple phenomenological model for localized dispersionless SF and generalized theory of SF [12] in the weak spin anharmonicity limit. Using the measured magnetic susceptibility of $\delta$-plutonium we show that its ground state is paramagnetic contrary to the predictions of the DFT calculations.

**Acknowledgements**

This work was supported by the State Atomic Energy Corporation of Russia "ROSATOM".

*E-mail address*: asolontsov@mail.ru